\documentclass[12pt,english,doc,apacite]{apa6}
\usepackage{geometry}
\usepackage[T1]{fontenc}
\usepackage[latin9]{inputenc}
\usepackage{array}
\usepackage{url}
\usepackage{bm}
\usepackage{multirow}
\usepackage{graphicx}
\usepackage{esint}
\usepackage{times}

\makeatletter

\providecommand{\tabularnewline}{\\}

\newenvironment{reflist}{ 
\begin{list}{}{\leftmargin=1em \itemindent=-1em}
}{
\end{list}
}
 
\makeatother

\usepackage{babel}
\begin{document}

\title{Bayesian meta-analysis of correlation coefficients through power prior }

\twoauthors{Zhiyong Zhang, Kaifeng Jiang, Haiyan Liu}{In-Sue Oh}

\twoaffiliations{University of Notre Dame}{Fox School of Business, Temple University}

\shorttitle{Bayesian meta-analysis}

\authornote{Correspondance should be sent to Zhiyong Zhang, 118 Haggar Hall, Univeristy of Notre Dame, IN 46556. Email: zzhang4@nd.edu.}
\abstract{To answer the call of introducing more Bayesian techniques to organizational
research (e.g., Kruschke, Aguinis, \& Joo, 2012; Zyphur \& Oswald,
2013), we propose a Bayesian approach for meta-analysis with power
prior in this article. The primary purpose of this method is to allow
meta-analytic researchers to control the contribution of each individual
study to an estimated overall effect size though power prior. This
is due to the consideration that not all studies included in a meta-analysis
should be viewed as equally reliable, and that by assigning more weights
to reliable studies with power prior, researchers may obtain an overall
effect size that reflects the population effect size more accurately.
We use the relationship between high-performance work systems and
financial performance as an example to illustrate how to apply this
method to organizational research. We also provide free online
software that can be used to conduct Bayesian meta-analysis proposed
in this study. Research implications and future directions are 
discussed.}

\maketitle
\setcounter{page}{1}

\section{INTRODUCTION}

Meta-analysis is a statistical method of combining findings from multiple
studies to get a more comprehensive understanding of the population
(Hunter \& Schmidt, 2004). A simple way to combine studies is to calculate
the weighted average of correlations between two variables (or differences
between two treatments) with the sample size being the weight (e.g.,
Hunter \& Schmidt, 2004). In addition to the averaged effect size,
the variation of it can also be investigated to see how the effect
size changes from one study to another. This method has become increasingly
popular in management research in recent years. According to a review
by Aguinis, Dalton, Bosco, Pierce, and Dalton (2011), thousands of
meta-analyses have been conducted and published in five major management
journals from 1982 to 2009. The number of annually published meta-analytically
derived effect sizes is also expected to keep growing in the future. 

Both fixed-effects and random-effects models have been used in meta-analysis
(e.g., Field, 2001; Hedges \& Olkin, 1985; Hedges \& Vevea, 1998;
Hunter \& Schmidt, 2004). Fixed-effects models assume the population
under study is fixed and homogenous and the finding from each study
provides an estimate, ideally unbiased or consistent, of the population
effect. For example, if one wants to study the relationship between
job satisfaction and job performance, he or she may believe the relationship
between the two variables is universal in the population. The differences
in the reported correlations in identified studies simply result from
sampling variation. Random-effects models assume the population is
variable and heterogeneous and can show different effects according
to the distinct features that characterize it. For example, for different
types of measures, research designs, and research samples, the relationship
between job satisfaction and job performance can be quite different
(Judge, Thoresen, Bono, \& Patton, 2001). Therefore, the differences
in the reported correlations reflect the heterogeneous effect sizes
in the population.

Meta-analysis has been conducted within both the frequentist and Bayesian
frameworks although arguably meta-analysis can naturally be viewed
as a Bayesian method in general. The frequentist methods for meta-analysis
can be found in many places such as Hedges and Olkin (1985), Hunter
and Schmidt (2004), and Rosenthal (1991). Relatively few studies have
discussed Bayesian meta-analysis (e.g., Carlin, 1992; Morris, 1992;
Smith, Spiegelhalter, \& Thomas, 1995; Steele \& Kammeyer-Mueller,
2008), which has been considered as having several advantages, such
as ``full allowance for all parameter uncertainty in the model, the
ability to include other pertinent information that would otherwise
be excluded, and the ability to extend the models to accommodate more
complex, but frequently occurring, scenarios'' (Sutton \& Abrams,
2001, p. 277). 

Traditional meta-analysis, using either the frequentist or Bayesian
approach, typically treats each study equivalently. In other words,
each study contributes equally to estimated overall effect size after
considering the weights proportional to sample sizes. However, in
many cases, not all studies included in a meta-analysis should make
equal contribution to the overall effect size; treating them equivalently
might cause unexpected consequences in meta-analysis. For example,
strategic management scholars may be interested in the relationships
between financial performance and its antecedents, such as human resource
management (HRM) practices (Combs, Liu, Hall, \& Ketchen, 2006) and
human capital (Crook, Todd, Combs, Woehr, \& Ketchen, 2011). Financial
performance can be measured objectively using data from archival data
or subjectively using survey data. Although both objective and subjective
measures are widely adopted in the literature, objective information
may reflect a firm's financial status more accurately than subjective
ratings because the latter involves more cognitively demanding assessments
and the informants may not always have the best knowledge of the information.
Therefore, those using objective measures may provide more reliable
information of the relationships between financial performance and
other variables than those based on subjective measures. For another
example, due to the difficulty of collecting longitudinal data, longitudinal
studies often result in a relatively small sample size compared with
cross-sectional studies obtaining all information from a single source.
Even though longitudinal designs may help avoid common method bias
and reduce inflation of correlations (Podsakoff, MacKenzie, Lee, \&
Podsakoff, 2003), their small sample sizes make them contribute less
to the final result. Instead, the cross-sectional studies with inflated
relationships may easily dominate the overall effect size because
of their large sample sizes. As illustrated in the two examples, treating
individual studies equivalently may produce potential misleading results.
However, not much attention has been paid to this issue when estimating
overall effect size in traditional meta-analysis. To address this
research need, this study proposes a Bayesian method for meta-analysis
that can control the contribution of each individual study through
power prior. As we discuss below, this method can allow meta-analysis
researchers more flexibility to estimate overall effect size by specifying
power parameters for individual studies. 

For illustration, the current study focuses on the meta-analysis of
sample correlation although the same method can be applied for other
effect size measures. In the following, we first demonstrate the use
of power prior through a fixed-effects model and then we extend our
method to random-effect models and meta-regression. Free online software
is introduced to carry out the Bayesian meta-analysis discussed in
this study. The use of Bayesian meta-analysis is further demonstrated
through a real meta-analysis example.

\section{BAYESIAN META-ANALYSIS THROUGH POWER PRIOR}

The proposed method is derived based on the Fisher z-transformation
of correlation. Suppose $\rho$ is the population correlation of two
variables that follow a bivariate normal distribution. For a given
sample correlation $r$ from a sample of $n$ independent subjects,
its Fisher z-transformation, denoted by $z$, is defined as
\[
z=\frac{1}{2}\ln\frac{1+r}{1-r}.
\]
$z$ approximately follows a normal distribution with mean 
\[
\frac{1}{2}\ln\frac{1+\rho}{1-\rho}
\]
and variance $\phi=\frac{1}{n-3}$ (Fisher, 1921).

Meta-analysis of correlation concerns the analysis of correlation
between two variables when a set of studies regarding the relationship
between the two variables are available. Suppose there are $m$ studies
that report the sample correlation between two variables. Each study
reports a sample correlation $r_{i}$ with the corresponding sample
size $n_{i}$. Let $z_{i}=\frac{1}{2}\ln\frac{1+r_{i}}{1-r_{i}}$
denote the Fisher z-transformation of $r_{i}$ and $\zeta_{i}=\frac{1}{2}\ln\frac{1+\rho_{i}}{1-\rho_{i}}$
be the Fisher z-transformation of the population correlation. Then,
$z_{i}\sim N(\zeta_{i},\phi_{i})$ with $\phi_{i}=(n_{i}-3)^{-1}$.

\subsection{Fixed-effects Models}

We first investigate the situation where the population can be considered
as homogeneous and, therefore, a fixed-effects model can be used.
In this case, the population correlation is
\[
\zeta_{i}\equiv\zeta=\frac{1}{2}\ln\frac{1+\rho}{1-\rho}
\]
and $z_{i}\sim N(\zeta,\phi_{i})$. 

The use of Bayesian methods requires the specification of priors (e.g.,
Gelman, Carlin, Stern, \& Rubin, 2003), which provides a perfect way
to conduct meta-analysis. A prior represents information on the population
correlation, or its Fisher z-transformation, without any data collection.
Although a prior is required, it may consist of ``no'' information
through certain types of prior such as Jeffreys' prior (e.g., Gill,
2002; Jeffreys, 1946). Suppose the prior for $\zeta$ follows a normal
distribution $N(\zeta_{0},\psi_{0})$ where $\zeta_{0}$ and $\psi_{0}$
are pre-determined values. For example, $\zeta$ could have a prior
N(0,1), which means a researcher initially believe the mean value
of $\zeta$ is 0, corresponding a correlation 0, with variance 1.
If little to none information is available, the so-called diffuse
prior can be used by specifying a large variance such as $\psi_{0}=10^{8}$. 

After collecting data, in the framework of meta-analysis, with the
availability of a study, one can get a better picture about the population
correlation. Bayesian methods provide a way to update the information
on the population correlation through Bayes' Theorem. Let $z_{1}$
denote the new information on the correlation after Fisher z-transformation
and $z_{1}\sim N(\zeta,\phi_{1})$. The distribution of the population
correlation $\zeta$ by combining the prior and the study is 
\begin{eqnarray*}
p(\zeta|z_{1}) & = & \frac{p(\zeta)p(z_{1}|\zeta)}{p(z_{1})},
\end{eqnarray*}
where $p(\zeta|z_{1})$ is called the posterior of $\zeta$ after
considering $z_{1}$. From Appendix A, we can conclude that the posterior
distribution is also a normal distribution $N(\zeta_{1},\psi_{1})$
where 
\begin{eqnarray}
\zeta_{1} & = & \frac{\frac{1}{\psi_{0}}\zeta_{0}+\frac{1}{\phi_{1}}z_{1}}{\frac{1}{\psi_{0}}+\frac{1}{\phi_{1}}}\label{eq:posterior_single}\\
\psi_{1} & = & \frac{1}{\frac{1}{\psi_{0}}+\frac{1}{\phi_{1}}}.
\end{eqnarray}
Therefore, the posterior mean $\zeta_{1}$ is the weighted average
of prior mean $\zeta_{0}$ and $z_{1}$ where the weights are the
inverse of variances of prior and data. If the prior is very accurate,
e.g., with a small variance, the prior mean will exert a big effect
on the posterior. For an extreme case, if $\psi_{0}=0$, the posterior
mean is $\zeta_{0}$, which is also the prior mean. On the other hand,
if only little prior information is available, reflected by large
variance of prior, the prior mean has little influence on the posterior.
For a special case where $\psi_{0}=+\infty$, the posterior mean is
$z_{1}$ and therefore, the posterior is fully determined by data.

The above analysis assumes that $z_{1}$ is fully reliable or the
researcher wants to utilize full information from $z_{1}$. However,
if, for practical reason, the information in $z_{1}$ is not accurate
enough (e.g., obtained from a flawed research design), it might distort
the posterior. In this situation, a researcher might prefer using
only partial information from $z_{1}$. Using the power prior idea
developed by Ibriham and Chen (2000), we can get the posterior
\begin{eqnarray}
p(\zeta|z_{1},\alpha_{1}) & = & \frac{p(\zeta)[p(z_{1}|\zeta)]^{\alpha_{1}}}{p(z_{1})},\label{eq:posteriorz1}
\end{eqnarray}
where $\alpha_{1}$ is a power parameter. Note that if $\alpha_{1}=0$,
no information from $z_{1}$ is used while when $\alpha_{1}=1$, full
information of $z_{1}$ is used. Partial information of $z_{1}$ can
be utilized by setting $\alpha_{1}$ to be a value between 0 and 1.
It can be shown (see Appendix B) that the posterior is still a normal
distribution with $N(\zeta_{1}^{*},\psi_{1}^{*})$ where
\begin{eqnarray*}
\zeta_{1}^{*} & = & \frac{\frac{1}{\psi_{0}}\zeta_{0}+\frac{\alpha_{1}}{\phi_{1}}z_{1}}{\frac{1}{\psi_{0}}+\frac{\alpha_{1}}{\phi_{1}}}\\
\psi_{1}^{*} & = & \frac{1}{\frac{1}{\psi_{0}}+\frac{\alpha_{1}}{\phi_{1}}}.
\end{eqnarray*}
Again the posterior mean is a weighted average of the prior mean and
$z_{1}$. However, note that the weight is different from the previous
situation because it is related to the power $\alpha_{1}$. If $\alpha_{1}<1$,
then the weight for $z_{1}$ is smaller than the previous results
in Equation \ref{eq:posterior_single}. This means posterior will
rely more on the prior. Note that this is equivalent to let $z_{1}\sim N(\zeta,\frac{\phi_{1}}{\alpha_{1}})$.
The information is passed through a normal distribution with the same
mean but enlarged variance.

Suppose without data collection, a researcher's prior information
on $\zeta$ is $N(0,1)$. One study in the literature reported a correlation
0.5 with the sample size 28 and, therefore, $z_{1}=0.549$ with variance
0.04. Table \ref{tbl:power} shows the posterior mean and variance
for $\zeta$ with power $\alpha_{1}$ ranges from 0 to 1. When $\alpha_{1}=0$,
the posterior is the same as the prior. When $\alpha_{1}$ increases
from 0.1 to 1, the posterior mean changes towards to $z_{1}$ because
more information from $z_{1}$ is included in the posterior. Furthermore,
the posterior variance is also becoming smaller. In summary, the use
of power $\alpha_{1}$ influences both the posterior mean and posterior
variance and can control the contribution of data to the posterior.

\begin{table}
\caption{The influence of the selection of power parameters for a single study}
\label{tbl:power}

\begin{tabular}{cccc}
\hline 
Data &  & z-transformation & Variance\tabularnewline
\hline 
$r_{1}=0.5$ &  & 0.549 & 0.04\tabularnewline
Prior &  & 0 & 1\tabularnewline
\hline 
Power &  & \multicolumn{2}{c}{Posterior}\tabularnewline
\cline{1-1} \cline{3-4} 
$\alpha_{1}$ &  & Mean & Variance\tabularnewline
0 &  & 0 & 1\tabularnewline
0.1 &  & 0.392 & 0.286\tabularnewline
0.2 &  & 0.458 & 0.167\tabularnewline
0.3 &  & 0.485 & 0.118\tabularnewline
0.4 &  & 0.499 & 0.091\tabularnewline
0.5 &  & 0.509 & 0.074\tabularnewline
0.6 &  & 0.515 & 0.063\tabularnewline
0.7 &  & 0.520 & 0.054\tabularnewline
0.8 &  & 0.523 & 0.048\tabularnewline
0.9 &  & 0.526 & 0.043\tabularnewline
1 &  & 0.528 & 0.038\tabularnewline
\hline 
\end{tabular}
\end{table}

In meta-analysis, data from multiple studies are available. Bayesian
methods provide a natural way to combine the data together. For example,
suppose we have another study with transformed correlation $z_{2}$
and its variance $\phi_{2}$ as well as the sample size $n_{2}$.
Furthermore, the power $\alpha_{2}$ is used when combining this study.
We have already obtained the posterior of $\zeta$ with the first
study in Equation \ref{eq:posteriorz1}. To get the posterior by combining
$z_{2}$, we can simply view the posterior in Equation \ref{eq:posteriorz1}
as a new prior. Then, the posterior of $\zeta$ with both $z_{1}$
and $z_{2}$ is
\[
p(\zeta|z_{1},z_{2},\alpha_{1},\alpha_{2})=\frac{p(\zeta|z_{1},\alpha_{1})[p(z_{2}|\zeta)]^{\alpha_{2}}}{p(z_{2})}.
\]
From Appendix C, we know posterior distribution is a normal distribution
$N(\zeta_{2}^{*},\psi_{2}^{*})$ where
\begin{eqnarray*}
\zeta_{2}^{*} & = & \frac{\frac{1}{\psi_{0}}\zeta_{0}+\frac{\alpha_{1}}{\phi_{1}}z_{1}+\frac{\alpha_{2}}{\phi_{2}}z_{2}}{\frac{1}{\psi_{0}}+\frac{\alpha_{1}}{\phi_{1}}+\frac{\alpha_{2}}{\phi_{2}}}\\
\psi_{2}^{*} & = & \frac{1}{\frac{1}{\psi_{0}}+\frac{\alpha_{1}}{\phi_{1}}+\frac{\alpha_{2}}{\phi_{2}}}.
\end{eqnarray*}
Clearly, the posterior mean is a weighted average of prior and the
two studies. More generally, if we have $m$ studies with $z_{i}$,
$n_{i}$, and $\alpha_{i}$, the posterior distribution of $\zeta$
is $N(\zeta_{m}^{*},\psi_{m}^{*})$ with 
\begin{eqnarray*}
\zeta_{m}^{*} & = & \frac{\frac{1}{\psi_{0}}\zeta_{0}+\sum_{i=1}^{m}\frac{a_{i}}{\phi_{i}}z_{i}}{\frac{1}{\psi_{0}}+\sum_{i=1}^{m}\frac{a_{i}}{\phi_{i}}}\\
\psi_{m}^{*} & = & \frac{1}{\frac{1}{\psi_{0}}+\sum_{i=1}^{m}\frac{a_{i}}{\phi_{i}}}.
\end{eqnarray*}

For illustration, we show the combination of two studies where the
first study reported a correlation 0.5 with the sample size 28 and
the second study reported a correlation 0 with the sample size 103.
Therefore, $z_{1}=0.549$ with variance 0.04 and $z_{2}=0$ with variance
0.01. A diffuse prior $N(0,100)$ is used here so that the effect
of prior is minimized. Table \ref{tbl:power-2study} presents the
posterior mean and variance for the population correlation with different
combinations of power for the two studies. First, when no information
from the two studies is utilized ($\alpha_{1}=\alpha_{2}=0$), the
posterior is just the prior. Second, when only the information of
Study 1 is fully used ($\alpha_{1}=1$, $\alpha_{2}=0$), the posterior
mean and variance are essentially the same as the Fisher z-transformation
and its variance of Study 1 because of the use of the diffuse prior.
Similarly, one can solely use the information from Study 2 by setting
$\alpha_{1}=0$ and $\alpha_{2}=1$. Third, when the information of
the two studies are used fully ($\alpha_{1}=\alpha_{2}=1$), the posterior
mean is about 0.110, the weighted average of 0.549 and 0 but leaning
towards 0 because the second study has a smaller variance. When setting
$\alpha_{1}=\alpha_{2}=0.5$, the posterior mean is still 0.110 but
the variance is about 0.016, twice of that when $\alpha_{1}=\alpha_{2}=1$.
This is because only partial information is used from the two studies.
Similar results can be seen from the table when other combination
of power is used. In summary, by controlling the power parameter,
one can control the contribution of each study to meta-analysis.

\begin{table}
\caption{The influence of the selection of power parameters for combining two
studies}
\label{tbl:power-2study}

\begin{tabular}{ccccc}
\hline 
\multicolumn{2}{c}{Data} &  & z-transformation & Variance\tabularnewline
\multicolumn{2}{c}{$r_{1}=0.5$} &  & 0.549 & 0.04\tabularnewline
\multicolumn{2}{c}{$r_{2}=0$} &  & 0 & .01\tabularnewline
\multicolumn{2}{c}{Prior} &  & 0 & 100\tabularnewline
\hline 
\multicolumn{2}{c}{Power} &  & \multicolumn{2}{c}{Posterior}\tabularnewline
\cline{1-2} \cline{4-5} 
$\alpha_{1}$ & $\alpha_{2}$ &  & Mean & Variance\tabularnewline
0 & 0 &  & 0 & 100\tabularnewline
1 & 0 &  & 0.549 & 0.040\tabularnewline
0 & 1 &  & 0.000 & 0.010\tabularnewline
0.1 & 1 &  & 0.013 & 0.010\tabularnewline
1 & 0.1 &  & 0.392 & 0.029\tabularnewline
0.5 & 0.5 &  & 0.110 & 0.016\tabularnewline
0.2 & 1 &  & 0.026 & 0.010\tabularnewline
1 & 0.2 &  & 0.305 & 0.022\tabularnewline
0.2 & 0.8 &  & 0.032 & 0.012\tabularnewline
0.8 & 0.2 &  & 0.275 & 0.025\tabularnewline
1 & 1 &  & 0.110 & 0.008\tabularnewline
\hline 
\end{tabular}
\end{table}

\subsection{Random-effects Models}

When the population is not homogeneous, it is not reasonable to assume
that $z_{i}$ has the same mean $\zeta$. Therefore, we discuss the
random-effects models in the Bayesian framework. A random-effects
model can be written as a two-level model,

\begin{equation}
\left\{ \begin{array}{l}
z_{i}=\zeta_{i}+e_{i}\\
\zeta_{i}=\zeta+v_{i}
\end{array}\right.
\end{equation}
where $Var(e_{i})=\phi_{i}$ and $Var(v_{i})=\tau$. In the model,
each $z_{i}$ has its mean $\zeta_{i}$ and the grand mean of $\zeta_{i}$
is $\zeta$. Based on Fisher z-transformation, $z_{i}\sim N(\zeta_{i},\phi_{i})$.
It is often assumed that $v_{i}$ has a normal distribution and, therefore,
$\zeta_{i}\sim N(\zeta,\tau)$. For the random-effects model, we have
the fixed-effects parameter $\zeta$ and the random-effects parameter
$\tau$. The parameter $\tau$ represents the between-study variability.
The parameter $\zeta$ can be transformed back to correlation that
represents the overall correlation across all studies. In addition,
we can also estimate the random effects $\zeta_{i}$, which can be
transformed back to correlations for individual studies. 

As for the fixed-effects models, to estimate model parameters for
the random-effects models, we need to specify priors. In this study,
the normal prior $N(\zeta_{0},\psi_{0})$ is used for $\zeta$ and
the inverse gamma prior $IG(\delta_{0},\gamma_{0})$ is used for $\tau$
with $\zeta_{0}$, $\psi_{0}$, $\delta_{0}$ and $\gamma_{0}$ denoting
known constants. In practice, $\zeta_{0}=0$, $\psi_{0}=10^{6}$,
$\delta_{0}=10^{-3}$ and $\gamma_{0}=10^{-3}$ are often used to
reduce the influence of priors. With the priors, the conditional posteriors
for $\zeta$, $\tau$, and $\zeta_{i}$ can be obtained as in Appendix
D. Then, the following Gibbs sampling procedure can be used to get
a Markov chain for each parameter.
\begin{APAenumerate}
\item Choose a set initial values for $\zeta$ and $\tau$, e.g., $\zeta^{(0)}=0$
and $\tau^{(0)}=1$.
\item Generate $\zeta_{i}^{(1)},i=1,\ldots,m$ from the normal distribution
\[
N\left(\frac{\frac{\zeta^{(0)}}{\tau^{(0)}}+\frac{z_{i}\alpha_{i}}{\phi_{i}}}{\frac{1}{\tau^{(0)}}+\frac{\alpha_{i}}{\phi_{i}}},\frac{1}{\frac{1}{\tau^{(0)}}+\frac{\alpha_{i}}{\phi_{i}}}\right).
\]

\item Generate $\tau^{(1)}$ from the inverse Gamma distribution IG($\delta_{0}+m/2,\gamma_{0}+[\sum_{i=1}^{m}(\zeta_{i}^{(1)}-\zeta^{(0)})^{2}]/2$).
\item Generate $\zeta^{(1)}$ from the normal distribution 
\[
N\left(\frac{\frac{\sum_{i=1}^{m}\zeta_{i}^{(1)}}{\tau^{(1)}}+\frac{\zeta_{0}}{\psi_{0}}}{\frac{m}{\tau^{(1)}}+\frac{1}{\psi_{0}}},\frac{1}{\frac{m}{\tau^{(1)}}+\frac{1}{\psi_{0}}}\right).
\]

\item Let $\zeta^{(0)}=\zeta^{(1)}$ and $\tau^{(0)}=\tau^{(1)}$ and repeat
Steps 2-4 to get $\zeta^{(2)}$, $\tau^{(2)}$ and $\zeta_{i}^{(2)},i=1,\ldots,m$.
\end{APAenumerate}
The above algorithm can be repeated for $R$ times to get a Markov
chain for $\zeta$, $\tau$, and $\zeta_{i}$. It can be shown that
the Markov chains converge to their marginal distributions after a
certain period and therefore can be used to infer on the parameters
(e.g., Gelman et al, 2003). The period for the Markov chains to converge
is called the burn-in period. Suppose the burn-in period is $k$.
Then the rest of the Markov chain from $(k+1)$th iteration to the
$R$th iteration can be used to get the mean and variance of $\zeta$,
$\tau$, and $\zeta_{i}$. Because a research is ultimately interested
in the correlation, we can also get the Markov chain for $\rho=\frac{\exp(2\zeta)-1}{\exp(2\zeta)+1}$
and for $\rho_{i}=\frac{\exp(2\zeta_{i})-1}{\exp(2\zeta_{i})+1}$.

To illustrate the influence of power parameters on the random-effects
meta-analysis, we consider a simple example with three studies that
report correlations 0.5, 0 and -0.5 with sample sizes 103, 28 and
103. The Fisher z-transformed data and their variances are given in
Table \ref{tbl:random}. Table \ref{tbl:random} also reports the
estimated overall correlation $\rho$ and individual correlation $\rho_{i},i=1,2,3$.
When the power is 1 for all three studies, the estimated $\rho$ is
approximately 0. Note that the estimated individual population correlations
for the first and third studies are smaller than the observed ones.
This is called ``shrinkage'' or ``multilevel averaging'' effect
of multilevel analysis (e.g., Greenland, 2000). The estimated random
effects are pulled towards the average effects. If, based on expert
opinions or other information, we suspect the reported negative correlation
could be because of unreliable study, we might assign it a different
weight. For example, if we give the third study a power 0.1, the estimated
overall correlation becomes 0.061. Furthermore, if we assign a power
0.01, the overall population becomes 0.215. Therefore, the effect
of the observed unreliable negative correlation can be controlled
through chosen power parameters.

\begin{table}
\caption{The influence of the use of power parameters on random-effects meta-analysis$^{1}$}
\label{tbl:random}

\begin{tabular}{cccccccc}
\hline 
\multicolumn{3}{c}{Data} &  & \multicolumn{2}{c}{z-transformation} & \multicolumn{2}{c}{Variance}\tabularnewline
\multicolumn{3}{c}{$r_{1}=0.5$} &  & \multicolumn{2}{c}{0.549} & \multicolumn{2}{c}{0.01}\tabularnewline
\multicolumn{3}{c}{$r_{2}=0$} &  & \multicolumn{2}{c}{0} & \multicolumn{2}{c}{0.04}\tabularnewline
\multicolumn{3}{c}{$r_{3}=-0.5$} &  & \multicolumn{2}{c}{-0.549} & \multicolumn{2}{c}{0.01}\tabularnewline
\multicolumn{3}{c}{Prior} &  & \multicolumn{2}{c}{0} & \multicolumn{2}{c}{100}\tabularnewline
\hline 
\multicolumn{3}{c}{Power} &  & \multicolumn{4}{c}{Posterior mean}\tabularnewline
\cline{1-3} \cline{5-8} 
$\alpha_{1}$ & $\alpha_{2}$ & $\alpha_{3}$ &  & $\rho$ & $\rho_{1}$ & $\rho_{2}$ & $\rho_{3}$\tabularnewline
1 & 1 & 1 &  & -0.002 & 0.482 & -0.001 & -0.482\tabularnewline
1 & 1 & 0.1 &  & 0.061 & 0.476 & 0.022 & -0.305\tabularnewline
1 & 1 & 0.01 &  & 0.215 & 0.469 & 0.099 & 0.099\tabularnewline
\hline 
\end{tabular}
\end{table}

\subsubsection{Selection between fixed-effects and random-effects models}

The choice of the fixed-effects or random-effects models is often
a subjective decision. However, we suggest using two methods to assist
such a decision. A random-effects model is only beneficial when there
is significant variation in the population effect sizes. Therefore,
for the first method, we can test whether the variance of $\zeta_{i}$,
$\tau$, is significant. If it is significant, it suggests that a
random-effects model is preferred. Otherwise, the fixed-effects model,
as a special case of random-effects models, can be used. Some scholars
have argued that the significance test may not have enough power to
detect variation in population values due to the small number of studies
(Hunter \& Schmidt, 2004). Therefore, we recommend another method
to compare the fixed-effects model and the random-effects model through
the deviance information criterion (DIC, Spiegelhalter et al., 2002).
If the fixed-effects model has the smaller DIC, it is preferred. Otherwise,
the random-effects model is better used.

\subsection{Meta-regression Models}

When a random-effects model is suggested, it often indicates possible
heterogeneity in the population. Therefore, predictors or covariates
can be identified to explain such a heterogeneity. Suppose a set of
$p$ covariates are available, denoted by $x_{1},x_{2},\ldots,x_{p}$.
Then, a meta-regression model can be constructed as below

\begin{equation}
\left\{ \begin{array}{l}
z_{i}=\zeta_{i}+e_{i}\\
\zeta_{i}=\beta_{1}+\beta_{2}x_{1i}+\cdots+\beta_{p+1}x_{pi}+v_{i}=\mathbf{x}_{i}\bm{\beta}+v_{i}
\end{array}\right.,
\end{equation}
where $\bm{\beta}=(\beta_{1},\beta_{2},\ldots,\beta_{p+1})'$, $\mathbf{x}_{i}=(1,x_{1i},x_{2i},\ldots,x_{pi})$,
and $v_{i}\sim N(0,\tau)$. If a coefficient $\beta_{i}$ is significant,
$x_{p}$ is a significant predictor that might be related to the heterogeneity
of the population correlation. 

To estimate $\bm{\beta}$ and $\phi$, we specify the multivariate
normal prior for $\bm{\beta}$ as $N(\bm{\zeta}_{0},\mathbf{\Psi}_{0})$
and the inverse Gamma prior $IG(\delta_{0},\gamma_{0})$ for $\tau$.
Typically, we use the following hyper-parameters for the priors: $\bm{\zeta}_{0}=\mathbf{0}_{(p+1)\times1}$,
$\mathbf{\Psi}_{0}=10^{6}\mathbf{I}$ with $\mathbf{I}$ denoting
a $(p+1)\times(p+1)$ identity matrix, and $\delta_{0}=\gamma_{0}=10^{-3}$.

With the prior, the conditional posteriors for $\bm{\beta}$, $\tau$,
and $\zeta_{i}$ can be obtained as shown in Appendix E. The conditional
posterior distribution of $\tau$ is an inverse Gamma distribution
$\tau|\bm{\beta},\zeta_{i}\sim IG(\delta_{0}+m/2,\gamma_{0}+\sum_{i=1}^{m}(\zeta_{i}-\mathbf{x}_{i}\bm{\beta})^{2}/2)$.
The conditional posterior distribution for $\beta$ is still a multivariate
normal distribution 
\[
N\Bigg(\bigg(\mathbf{\Psi}_{0}^{-1}+\frac{\mathbf{X}'\mathbf{X}}{\tau}\bigg)^{-1}\bigg(\mathbf{\Psi}_{0}^{-1}\bm{\zeta}_{0}+\frac{\mathbf{X}'\mathbf{X}}{\tau}\hat{\bm{\beta}}\bigg),(\mathbf{\Psi}_{0}^{-1}+\frac{\mathbf{X}'\mathbf{X}}{\tau})^{-1}\Bigg)
\]
where $\hat{\bm{\beta}}$ is the least square estimate of $\bm{\beta}$
such that $\hat{\bm{\beta}}=(\mathbf{X}'\mathbf{X})^{-1}\mathbf{X}'\bm{\zeta}$
with $\mathbf{X}=(\mathbf{x}_{1},\ldots,\mathbf{x}_{m})'$ as the
design matrix and $\bm{\zeta}=(\zeta_{1},\zeta_{2},\ldots,\zeta_{m})'$.
The conditional posterior for $\zeta_{i}$ is 
\[
N\left(\frac{\frac{\alpha_{i}z_{i}}{\phi_{i}}+\frac{X_{i}\bm{\beta}}{\tau}}{\frac{\alpha_{i}}{\phi_{i}}+\frac{1}{\tau}},\frac{1}{\frac{\alpha_{i}}{\phi_{i}}+\frac{1}{\tau}}\right).
\]
With the set of conditional posteriors, the Gibbs sampling algorithm
can be used to generate Markov chain for each unknown parameter as
for the random-effects meta-analysis.

\section{SOFTWARE}

\begin{figure}
\caption{The interface of the online software metacorr }
\label{fig:interface}

\includegraphics[scale=0.9]{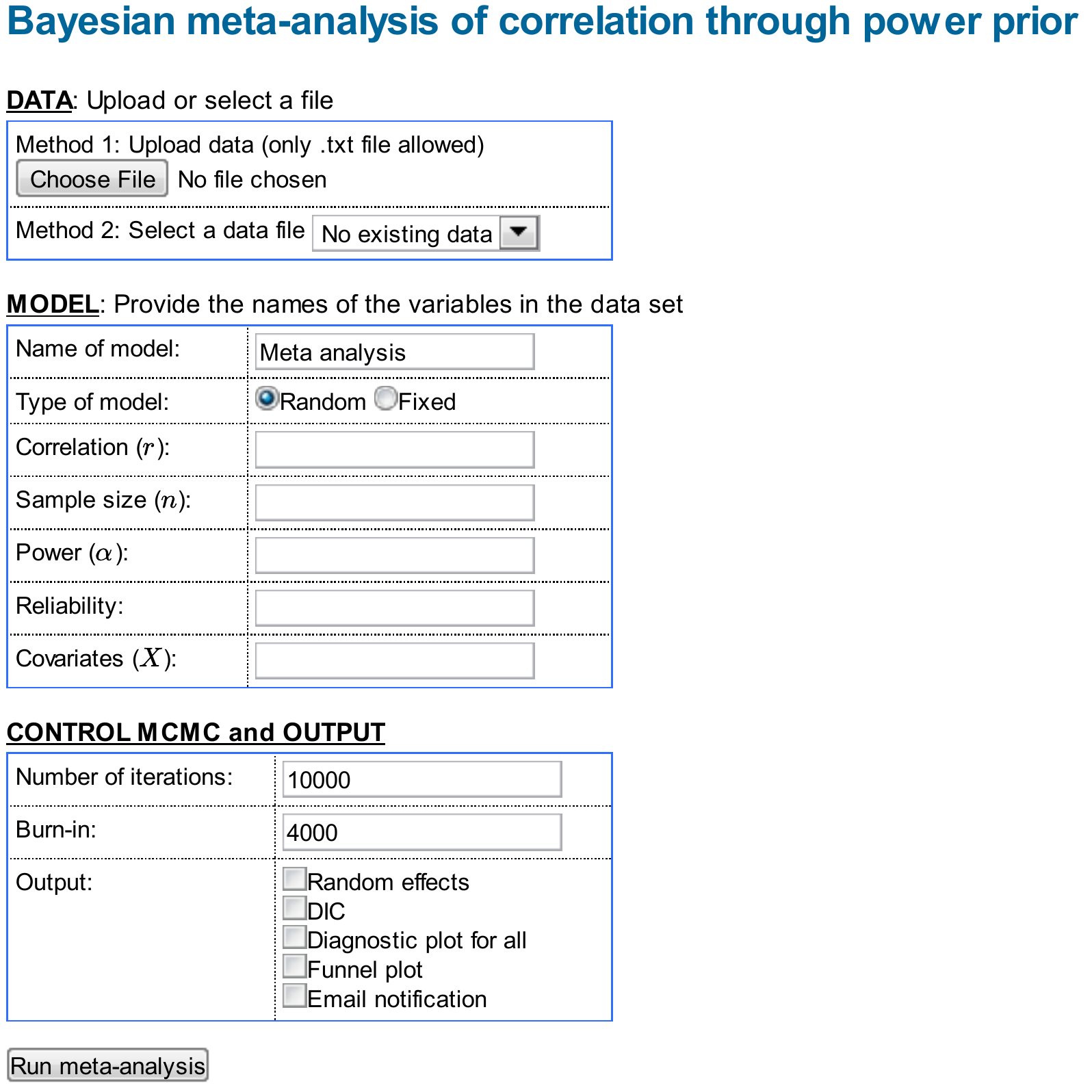}
\end{figure}

To facilitate the use of Bayesian meta-analysis method through power
prior, we developed a free online program that can be accessed with
the URL \url{http://webbugs.psychstat.org/modules/metacorr/}. The
online program can be used within a typical Web browser. It has an
interface shown in Figure \ref{fig:interface}. To use the program,
one needs either to upload a new data file or select an existing file.
Note names of the existing files are shown in the drop down menu.
The existing file has to be a text file in which the data values are
separated by one or more white spaces. The first line of the data
file will be the variable names, which will be used in the model.

Next, a user chooses the model to use. For example, the user can choose
to use either the random-effects model (default option) or the fixed-effects
model. Then, information on the model can be provided. Both the \emph{Correlation}
and \emph{Sample size} are required for all analysis, which can be
specified using the variable names in the data set. For example, if
we use ``fi'' to represent the correlation between financial performance
and another variable in the data set, then ``fi'' should be input
in the field of \emph{Correlation} in the interface. Similarly, ``n''
is used in the Sample size field because in the data set, ``n''
is also the variable name for sample size. In addition, a user can
also specify the variables for power and covariates used in the model. 

Finally, one can elect to control Markov chain Monte Carlo (MCMC)
method and output of the meta-analysis. For example, the total number
of Monte Carlo iteration and the burn-in period can be specified.
In the output, one can require the output of the estimates for random
effects $\zeta_{i}$, DIC, and diagnostic plots for all model parameters
including the random effects. If one checks the option \emph{Email
notification}, an email will be sent to the user once the analysis
is completed.

\section{AN EXAMPLE}

We use the relationship between high-performance work systems (HPWS)
and financial performance as an example to illustrate the use of Bayesian
meta-analysis with power prior. HPWS refers to a bundle of human resource
management (HRM) practices that are intended to enhance employees'
abilities, motivation, and opportunity to make contribution to organizational
effectiveness, including practices such as selective hiring, extensive
training, internal promotion, developmental performance appraisal,
performance-based compensation, flexible job design, and participation
in decision making (Lepak, Liao, Chung, \& Harden, 2006). Strategic
HRM scholars have devoted considerable effort to studying the influence
of HPWS on firm performance in the past three decades and consistently
found that the use of HPWS is positively related to employee and firm
performance (Paauwe, Wright, \& Guest, 2013). Indeed, recent meta-analyses
have demonstrated the positive relationships between HPWS and a variety
of performance outcomes (Combs et al., 2006; Jiang, Lepak, Hu, \&
Baer, 2012; Subramony, 2009), including employee outcomes (e.g., human
capital, employee motivation), operational outcomes (e.g., productivity,
service quality, and innovation), and financial outcomes (e.g., profit,
return on assets, and sales growth). The purpose of this study is
not to compare the results obtained from Bayesian meta-analysis to
those of previous research. Instead, we just use the research on HPWS
as an example and specifically focus on the relationship between HPWS
and financial performance, which is one of the most important considerations
of strategic HRM research. 

We used several search techniques to identify previous empirical studies
that examined the relationship between HPWS and financial performance.
First, we searched for published studies in the databases \emph{Business
Source Premier}, \emph{Google Scholar}, \emph{Web of Science}, and
\emph{PsycINFO} by combining keywords associated with HPWS (e.g.,
HRM systems, high-performance work systems, high-commitment HRM practices,
and high-involvement HRM practices) and with keywords related to financial
performance (e.g., financial performance, profit, return on assets,
return on equity, Tobin\textquoteright{}s Q, and sales growth). Second,
we checked the references of previous reviews on HPWS (e.g., Combs
et al., 2006; Jiang et al., 2012; Subramony, 2009) and added the articles
that were missed in the database searches. Third, we conducted a manual
search of major management journals that often publish strategic HRM
research, including \emph{Academy of Management Journal, Journal of
Management, Journal of Applied Psychology, Personnel Psychology, Strategic
Management Journal, Organization Science, Journal of Management Studies,
Human Resource Management, Human Resource Management Journal, and
International Journal of Human Resource Management}, to locate studies
that were not included in the previous searches. Finally, we searched
ProQuest Digital Dissertations and conference proceedings for the
annual meetings of the Academy of Management for unpublished dissertations
and conference papers from 2008 to 2013. 

We used four criteria to include studies in the following meta-analysis.
First, we only included studies that examined the relationship between
HPWS and financial performance at the unit-level of analysis (e.g.,
teams, stores, business units, and firms). Studies conducted at the
individual level of analysis were excluded from the analysis. Second,
we only included studies that examined HRM practices as a system.
Those examining the relationships between individual HRM practices
and financial performance were not considered in this research. Third,
consistent with previous meta-analyses (e.g., Combs et al., 2006;
Jiang et al., 2012; Subramony, 2009), we limited financial performance
to variables indicating financial or accounting outcomes, such as
profit, return on assets, return on invested capital, return on equity,
Tobin\textquoteright{}s Q, sales growth, and perceived financial performance.
Those only reporting the relationships between HPWS and other types
of outcomes (e.g., employee outcomes and operational outcomes) were
not included in the analysis. Fourth, the studies need to report at
least two pieces of information in order to be included in the meta-analysis
\textendash{} correlation coefficient between HPWS and financial performance
and sample size. This procedure resulted in 56 independent studies
that were entered in the following analysis. 

Before conducting Bayesian meta-analysis, we first corrected the observed
correlation from each sample for unreliability by following the procedure
outlined by Hunter and Schmidt (2004). Because HPWS has been considered
as a formative construct (Delery, 1998) for which a high internal
reliability (e.g., Cronbach\textquoteright{}s alpha) is not required,
we used a reliability of 1 for the measure of HPWS. Similarly, we
used a reliability of 1 for the objective measures of financial performance
and used Cronbach\textquoteright{}s alpha as the reliability of the
subjective measures of financial performance. 

In addition, we consider firm size as a potential moderator of the
relationship between HPWS and financial performance in order to test
the meta-regression model of this study. Firm size is commonly included
as a control variable in strategic HRM research, but its moderating
effect has rarely been explored in either primary studies or a meta-analysis.
Two competing hypotheses can be proposed in terms of its moderating
role. On the one hand, some researchers have suggested that large
organizations are likley to use more sophisticated HRM practices (e.g.,
HPWS) compared with small and mediem enterprises (e.g., Guthrie, 2001;
Jackson \& Schuler, 1995). As firm size increases, firms may also
have more advantages such as economy of scale (e.g., Pfeffer \& Salancik,
1978) and thus be more likely to gain benefit from their investment
in HRM practices. On the other hand, large firms' financial performance
may be more affected by other factors beyond human resources (Capon,
Farley, \& Hoenig, 1990). In this case, the role of HPWS in enhancing
financial performance may be limited in large firms than in small
and medium firms. Taking these considerations together, we expect
that firm size may moderate the relationship between HPWS and financial
performance but make no directional prediction of this effect. Firm
size is usually indicated by the number of employees. Studies with
average number of employees greater than 250 were coded as 1 (i.e.,
large firms) and the others were coded as 0 (i.e., small and medium
firms). 

Table \ref{tbl:summary} shows the summary statistics of the data
used in this example. Among the total of 56 studies, 46 measured financial
performance using the archival data (i.e., objective performance)
and 10 used subjective measures of financial performance (i.e., subjective
performance). In addition, 37 studies were coded as large firms and
19 were coded as small and medium firms. The observed correlations
ranged from 0.01 to 0.52 with sample sizes ranging from 50 to 2136.

\begin{table}
\caption{Summary statistics}
\label{tbl:summary}

\begin{tabular}{cccccc}
\hline 
 & Minmum & Mean & Median & Maximum & Standard deviation\tabularnewline
\hline 
Correlation & 0.01 & 0.22 & 0.200 & 0.52 & 0.13\tabularnewline
Sample size & 50 & 281 & 191 & 2136 & 325\tabularnewline
Reliability & 0.74 & 0.97 & 1 & 1 & 0.07\tabularnewline
 & \multicolumn{3}{c}{Small \& Medium: 19} & \multicolumn{2}{c}{Large: 37}\tabularnewline
 & \multicolumn{3}{c}{Objective studies: 46} & \multicolumn{2}{c}{Subjective studies: 10}\tabularnewline
\hline 
\end{tabular}
\end{table}

Four power schemes are considered in the meta-analysis. First, every
study is given the power of 1. In this case, every study contributes
to the meta-analysis result fully and equally. This is equivalent
to conduct traditional meta-analysis using Bayesian methods. Second,
the reliability of financial performance of each study is used as
power. The reason for this choice is that, if a measure is not reliable,
only partial information will be used in meta-analysis. Third, two
studies have sample sizes larger than 1000 (1212 and 2136, respectively).
In order to avoid the dominant influence of the two studies on the
final result, we assign them a power of 0.1 and the rest of studies
a power of 1 in meta-analysis. Fourth, arguably a study with a large
effect size is more likely to be published, which might cause publication
bias. Therefore, reducing the influence of the studies with large
effect sizes might be helpful in reducing publication bias. In this
power scheme, we set the power at 0.5 for studies with correlations
larger than 0.2. For the power schemes 3 and 4, the choice of power
is rather liberal. A more serious analysis might consider different
levels of power.

\subsection{Results of Fixed-effects Meta-analysis}

We first apply the fixed-effects meta-analysis model to the example
data. Table \ref{tbl:fixed results} shows the results using the four
different power schemes. When every study is assigned the equal power
of 1 (Power 1), the estimated population correlation $\rho$ is 0.263
($\zeta$ is the Fisher z-transformed estimate). If the reliability
of financial performance is used as power (Power 2), the estimated
correlation is about 0.264. However, when the two studies with the
largest sample size are assigned a power of 0.5 (Power 3), the estimated
correlation becomes 0.226. Note the estimated correlation in this
condition is significantly different from the other two correlation
estimates simply based on the credible interval estimates. In the
observed studies, the correlations for the two study are 0.34 and
0.45, respectively, both of which are larger than the estimated fixed-effect
correlation. When no power is used, the two studies pull the estimates
close to them because their large sample sizes lead to big weights
in the estimated correlation. Under the situation where the studies
with large correlation are assigned a weight 0.5 (Power 4), the estimated
correlation is 0.22, which is even smaller than the situation of Power
3. This is because the large correlations are downweighted.

\begin{table}
\caption{Results from fixed-effects meta-analysis$^{2,3}$}
\label{tbl:fixed results}

\begin{tabular}{ccccccc}
\hline 
 &  & Estimate & sd & \multicolumn{2}{c}{CI} & DIC\tabularnewline
\hline 
\multirow{2}{*}{Power 1} & $\zeta$ & 0.27{*} & 0.008 & 0.254 & 0.285 & \multirow{2}{*}{184}\tabularnewline
 & $\rho$ & 0.26{*} & 0.007 & 0.249 & 0.278 & \tabularnewline
\hline 
\multirow{2}{*}{Power 2} & $\zeta$ & 0.27{*} & 0.008 & 0.255 & 0.286 & \multirow{2}{*}{175.9}\tabularnewline
 & $\rho$ & 0.26{*} & 0.008 & 0.249 & 0.279 & \tabularnewline
\hline 
\multirow{2}{*}{Power 3} & $\zeta$ & 0.23{*} & 0.009 & 0.212 & 0.247 & \multirow{2}{*}{72.11}\tabularnewline
 & $\rho$ & 0.23{*} & 0.008 & 0.209 & 0.242 & \tabularnewline
\hline 
\multirow{2}{*}{Power 4} & $\zeta$ & 0.22{*} & 0.009 & 0.205 & 0.242 & \multirow{2}{*}{77.44}\tabularnewline
 & $\rho$ & 0.22{*} & 0.009 & 0.202 & 0.237 & \tabularnewline
\hline 
\end{tabular}
\end{table}

\subsection{Results of Random-effects Meta-analysis}

Table \ref{tbl:random-results} shows the results from the random-effects
meta-analysis. First, the estimated correlations from the random-effects
and fixed-effects methods are quite different (0.23 vs. 0.27) when
the power is not considered. This is because for the random-effects
method, the between-study variability is considered. Therefore, extreme
studies (e.g., those with unusual large sample sizes) are shrunk towards
the average. Furthermore, within the random-effects method, there
is not much difference in the estimated correlation. Second, only
for power scheme 4, the estimated correlation shows notable difference
from the rest of the power schemes. The reason is because studies
with large correlations are downweighted. Third, in all situation,
the variance estimate of $\tau$ is significant. This indicates there
is sufficient variability in the studies to consider a random-effects
meta-analysis to model the heterogeneity in the population.

\begin{table}
\caption{Results from random-effects meta-analysis$^{2,3}$}
\label{tbl:random-results}

\centering{}%
\begin{tabular}{ccccccc}
\hline 
 &  & Estimate & sd & \multicolumn{2}{c}{CI} & DIC\tabularnewline
\hline 
\multirow{3}{*}{Power 1} & $\zeta$ & 0.23{*} & 0.02 & 0.191 & 0.269 & \multirow{3}{*}{-98.45}\tabularnewline
 & $\tau$ & 0.016{*} & 0.004 & 0.01 & 0.026 & \tabularnewline
 & $\rho$ & 0.226{*} & 0.019 & 0.189 & 0.263 & \tabularnewline
\hline 
\multirow{3}{*}{Power 2} & $\zeta$ & 0.23{*} & 0.02 & 0.191 & 0.27 & \multirow{3}{*}{-97.18}\tabularnewline
 & $\tau$ & 0.016{*} & 0.004 & 0.01 & 0.026 & \tabularnewline
 & $\rho$ & 0.226{*} & 0.019 & 0.189 & 0.263 & \tabularnewline
\hline 
\multirow{3}{*}{Power 3} & $\zeta$ & 0.228{*} & 0.02 & 0.19 & 0.267 & \multirow{3}{*}{-93.99}\tabularnewline
 & $\tau$ & 0.016{*} & 0.004 & 0.009 & 0.025 & \tabularnewline
 & $\rho$ & 0.224{*} & 0.019 & 0.187 & 0.261 & \tabularnewline
\hline 
\multirow{3}{*}{Power 4} & $\zeta$ & 0.218{*} & 0.02 & 0.178 & 0.259 & \multirow{3}{*}{-85.4}\tabularnewline
 & $\tau$ & 0.015{*} & 0.004 & 0.008 & 0.024 & \tabularnewline
 & $\rho$ & 0.214{*} & 0.019 & 0.177 & 0.253 & \tabularnewline
\hline 
\end{tabular}
\end{table}

\subsection{Results of Meta-regression}

From the random-effects meta-analysis, we concluded that the population
should be considered as heterogeneous.Through meta-regression analysis,
we investigate whether the heterogeneity is related to firm size of
different studies. Based on the results in Table \ref{tbl:meta-regression},
firm size is not significantly related to the individual differences
in the population correlation because the slope parameter $\beta_{2}$
is not significant regardless of the choice of power. Furthermore,
the results from the first three power schemes are very close. Comparing
all four power schemes, power scheme 4 has smaller intercept while
larger absolute slope. Combined, the results do not suggest the moderating
effect of firm size on the relationship between HPWS and financial
performance. It implies that HPWS used in both large firms and small
and medium firms are salutary for enhancing financial performance.

\begin{table}
\caption{Results from meta-regression$^{2,3}$}
\label{tbl:meta-regression}

\begin{tabular}{ccccccc}
\hline 
 &  & Estimate & sd & \multicolumn{2}{c}{CI} & DIC\tabularnewline
\hline 
\multirow{3}{*}{Power 1} & $\beta_{1}$(intercept) & 0.248{*} & 0.034 & 0.181 & 0.316 & \multirow{3}{*}{-97.9}\tabularnewline
 & $\beta_{2}$(size) & -0.028 & 0.042 & -0.113 & 0.053 & \tabularnewline
 & $\tau$ & 0.017{*} & 0.004 & 0.01 & 0.026 & \tabularnewline
\hline 
\multirow{3}{*}{Power 2} & $\beta_{1}$(intercept) & 0.249{*} & 0.034 & 0.181 & 0.317 & \multirow{3}{*}{-96.63}\tabularnewline
 & $\beta_{2}$(size) & -0.029 & 0.042 & -0.113 & 0.053 & \tabularnewline
 & $\tau$ & 0.016{*} & 0.004 & 0.01 & 0.026 & \tabularnewline
\hline 
\multirow{3}{*}{Power 3} & $\beta_{1}$(intercept) & 0.245{*} & 0.034 & 0.179 & 0.312 & \multirow{3}{*}{-93.5}\tabularnewline
 & $\beta_{2}$(size) & -0.027 & 0.042 & -0.111 & 0.054 & \tabularnewline
 & $\tau$ & 0.016{*} & 0.004 & 0.009 & 0.025 & \tabularnewline
\hline 
 & $\beta_{1}$(intercept) & 0.24{*} & 0.035 & 0.172 & 0.31 & \multirow{3}{*}{-84.79}\tabularnewline
Power 4 & $\beta_{2}$(size) & -0.034 & 0.043 & -0.121 & 0.048 & \tabularnewline
 & $\tau$ & 0.015{*} & 0.004 & 0.008 & 0.025 & \tabularnewline
\hline 
\end{tabular}
\end{table}

\subsection{Fixed-effects Meta-analysis, Random-effects Meta-analysis, or Meta-regression?}

Selection among different methods deserves much more investigation.
Here, we just illustrate several possibilities using the example above,
which certainly have their limitations. In choosing between fixed-effects
and random-effects meta-analysis, one can check whether the variance
parameter $\tau$ from the random-effects meta-analysis is significant.
If it is significant, random-effects meta-analysis can be used. Our
example showed that random-effects meta-analysis might be preferred.
For the choice between random-effects meta-analysis and meta-regression,
one can focus on the significance of the regression coefficients for
predictors. If the coefficients are not significant, it might suggest
that there is no need to include the proposed predictor in meta-regression. 

We can also directly compare fixed-effects meta-analysis, random-effects
meta-analysis, and meta-regression using DIC. The model with the smallest
DIC indicates it fits the data best. However, DIC should only be used
to compare models under the same power scheme. The calculation of
DIC across power schemes would utilize different information and therefore
is not valid. For example, under power scheme 1, DICs for the three
models are 184, -98.45, and -97.9. This suggests that the meta-regression
fits the current data best. However, there is no given cut-off on
when a model can be considered as fitting data significantly better.

\section{DISCUSSION}

The current study presents a Bayesian method for meta-analysis. A
unique feature of our method is to enable researchers to control the
contribution of individual studies included in a meta-analysis through
power prior. The motivation of this approach comes from the notion
that not all studies should be treated equivalently when estimating
the overall effect size in a meta-analysis. By developing an online
program and using the example of the relationship between HPWS and
financial performance, we have shown how to apply this method into
management research. In the rest of this article, we briefly summarize
the example results derived from the method we proposed. And then
we discuss some implications of this method to meta-analysis in the
field of management. 

In the example study, we use four power schemes to assign powers to
individual studies included in the meta-analysis. As shown in fixed-effects,
random-effects, and meta-regression models, using the reliability
of financial performance as power does not dramatically change the
results obtained from regular meta-analysis that uses full information
provided by each study. This is because that only ten studies used
subjective measures of financial performance and the use of reliability
as power would only influence how the ten out of 56 studies contribute
to the final results. Moreover, the reliabilities for the subjective
measures are typically high, so the vast majority of the information
they provide still contributes to the overall effect size. If one
uses another example with more subjective measures, the difference
in effect size between regular meta-analysis and meta-analysis using
reliability as power may be more obvious. Either way, our method provides
a way to evaluate whether reliability influences meta-analysis results.

When power is used to reduce the influence of two studies with large
sample sizes, the overall effect size in fixed-effects model becomes
significantly different from what is obtained in the regular model,
and the change is less obvious in random-effects and meta-regression
models. This is because between-study variability is taken into account
in random-effects model, which can shrink extreme effect sizes towards
the average. However, this does not mean that using power to modify
the impact of extremely large samples always has a larger impact on
fixed-effects model than on random-effects model. It may also depend
on the observed correlations of studies with large sample sizes. For
example, if the correlation of a large sample is similar to the weighted
average of the rest of the studies, assigning a small power to the
large sample may not significantly change the overall effect size
in either fixed-effects model or random-effects model. 

The influence of power becomes more salient under power scheme 4 where
studies with correlations larger than 0.2 are assigned a power of
0.5. We argue that this setting can potentially be used to deal with
publication bias. For example, if we believe the studies with large
effect sizes are over-sampled, we can assign them power smaller than
1. On the other hand, if one believes the studies with small effect
sizes are under-sampled, power larger than 1 can also be used. Certainly
the choice of power needs careful consideration.

Combined, the example of the relationship between HPWS and financial
performance provides an initial illustration of our Bayesian approach
of meta-analysis. We encourage researchers who are interested in this
approach to test their data using the developed software and compare
the results derived from different power schemes. In the following,
we shift attention to the implications of this method to meta-analysis
in management research. 

First of all, we want to make it clear that it is not necessary to
apply power prior to all meta-analyses. However, if researchers believe
certain factors may impact the credibility of research findings and
they can distinguish between more reliable and less reliable studies,
we would recommend them to estimate the overall effect size with power
prior, at least for the purpose of comparison. For example, reliability
has been commonly used to correct for measurement error in meta-analysis
(Hunter \& Schmidt, 2004). As reliability decreases, the corrected
effect size is more likely to be enlarged. However, a low reliability
often indicates poor quality of a measure, which may not accurately
reflect the intended construct. Therefore, the study using the measure
with low reliability may be less likely to represent the true relationship
of interest. In addition, research design (e.g., cross-sectional vs.
longitudinal, single-source data vs. multiple-source data, self-report
ratings vs. observer ratings, field studies vs. experimental studies)
may also affect the extent to which a study can provide reliable information
of the relationship between two variables. Combining study findings
without considering the credibility of the information may lead to
misleading results. Although researchers can summarize studies of
different characteristics separately (e.g., Judge, Colbert, \& Ilies,
2004; Oh, Wang, \& Mount, 2011; Sin, Nahrgang, \& Morgeson, 2009),
they may still need to combine all studies to yield overall effect
sizes that can be used as inputs of other analyses, such as meta-analytic
structural equation modeling (MASEM; Viswesvaran \& Ones, 1995). In
these cases, researchers may choose to use more information from reliable
studies through power prior in order to obtain overall effect sizes
that can better represent the true relationships. 

Bayesian meta-analysis with power prior can also be used to deal with
outliers, including outliers of observed correlations and outliers
of sample sizes. Traditionally, researchers often eliminate the most
extreme data points to attenuate the influence of outliers on overall
effect sizes (e.g., Hedges, 1987, Huber, 1980, Tukey, 1960). This
is similar to assigning a power of 0 to studies considered as outliers
and using no information of the eliminated studies in analysis. However,
rather than deleting the data points completely, researchers can also
choose to use only a small part of their information by assigning
a small non-zero power to those studies. 

One important issue that is out of the discussion of this article
is what power value should be assigned to each study in meta-analysis
with power prior. The method proposed in this study cannot determine
whether a power prior scheme is realistic or not to reflect the contribution
of each study to the final results. It is more reasonable for researchers
who are familiar with the nature of the included studies to make the
decisions. The general guideline is to identify the criteria that
can indicate the credibility of research findings and use it to guide
power prior decision in meta-analysis. One attempt of this study is
to use reliability as a power for studies relying on subjective measures,
which may reduce the overcorrection for unreliability due to extremely
low reliabilities. In addition, we recommend that one should always
compare the results from the analysis with and without power priors
to inform the influence of the use of power priors. We encourage more
efforts to further explore this issue in the future. 

This study can be improved and extended in many ways. First, in both
random-effects meta-analysis and meta-regression, we assume that the
random effects follow a normal distribution. This assumption might
not be valid when there are extreme values. Further study can incorporate
robust Bayesian analysis to deal with the problem (e.g., Zhang, Lai,
Lu, \& Tong, 2013). Second, the current study has focused on the development
of the method for correlation. However, the method can be equally
applied to other effect sizes such as mean differences and odds ratios.
Third, we have discussed the use of DIC for model comparison. The
performance of it can be investigate through rigor simulation in the
future.

\section{REFERENCES}
\begin{reflist}
\item Aguinis, H., Dalton, D. R., Bosco, F. A., Pierce, C. A., \& Dalton,
C. M. 2011. Meta-analytic choices and judgment calls: Implications
for theory building and testing, obtained effect sizes, and scholarly
impact. \emph{Journal of Management}, 37: 5-38.
\item Carlin, J. B. 1992. Meta-analysis for 2$\times$2 tables: A Bayesian
approach. \emph{Statistics in Medicine, }11: 141-158. 
\item Capon, N., Farley, J. U., \& Hoenig, S. 1990. Determinants of financial
performance: A meta-analysis. \emph{Management Science}, 36: 1143-1159.
\item Combs, J., Liu, Y., Hall, A., \& Ketchen, D. 2006. How much do high-performance
work practices matter? A meta-analysis of their effects on organizational
performance. \emph{Personnel Psychology}, 59: 501-528. 
\item Crook, T. R., Todd, S. Y., Combs, J. G., Woehr, D. J., \& Ketchen
Jr, D. J. 2011. Does human capital matter? A meta-analysis of the
relationship between human capital and firm performance. \emph{Journal
of Applied Psychology}, 96: 443-456. 
\item Delery, J. E. 1998. Issues of fit in strategic human resource management:
Implications for research. \emph{Human Resource Management Review},
8: 289-309. 
\item Field, A. P. 2001. Meta-analysis of correlation coefficients: A Monte
Carlo comparison of fixed- and random-effects methods. \emph{Psychological
Methods, }6: 161-180. 
\item Field, A. P., \& Gillett, R. 2010, How to do a meta-analysis. \emph{British
Journal of Mathematical and Statistical Psychology, }63: 665-694.
\item Fisher, R. A. 1921. On the probable error of a coefficient of correlation
deduced from a small sample. \emph{Metron}, 1: 3-32.
\item Gelman, A., Carlin, J. B., Stern, H. S., \& Rubin, D. B. 2003. \emph{Bayesian
data analysis}. Chapman \& Hall/CRC.
\item Gill, J. 2002. \emph{Bayesian methods: A social and behavioral sciences
approach}. CRC Press.
\item Greenland, S. 2000. Principles of multilevel modelling. \emph{International
Journal of Epidemiology, }29: 158-167.
\item Guthrie, J. P. 2001. High-involvement work practices, turnover, and
productivity: Evidence from New Zealand. \emph{Academy of management
Journal}, 44: 180-190.
\item Hedges, L. V. 1987. How hard is hard science, how soft is soft science:
The empirical cumulativeness of research. \emph{American Psychologist},
42: 443-455.
\item Hedges, L. V. (1992). Meta-analysis. \emph{Journal of Educational
Statistics}, 17: 279-296.
\item Hedges, L. V., \& Olkin, I. 1985. \emph{Statistical methods for meta-analysis}.
New York: Academic. 
\item Hedges, L. V., \& Vevea, J. L. 1998. Fixed- and random-effects models
in meta-analysis. \emph{Psychological Methods}, 3(4): 486-504. 
\item Huber, P. J. 1980. \emph{Robust statistics}. New York: Wiley. 
\item Hunter, J. E., \& Schmidt, F. L. 2004. \emph{Methods of meta-analysis:
Correcting error and bias in research findings} (2nd ed.). Newbury
Park, CA: Sage.
\item Ibrahim, J. G., \& Chen, M. H. 2000. Power prior distributions for
regression models. \emph{Statistical Science}, 15: 46-60. 
\item Jackson, S. E., \& Schuler, R. S. 1995. Understanding human resource
management in the context of organizations and their environments.
\emph{Annual Review of Psychology}, 46: 237-264. 
\item Jeffreys, H. 1946. An invariant form for the prior probability in
estimation problems. \emph{Proceedings of the Royal Society of London.
Series A, Mathematical and Physical Sciences, }186(1007): 453-461.
\item Jiang, K., Lepak, D. P., Hu, J., \& Baer, J. C. 2012. How does human
resource management influence organizational outcomes? A meta-analytic
investigation of mediating mechanisms. \emph{Academy of Management
Journal}, 55: 1264-1294.
\item Judge, T. A., Colbert, A. E., \& Ilies, R. 2004. Intelligence and
leadership: a quantitative review and test of theoretical propositions.
\emph{Journal of Applied Psychology}, 89: 542-552. 
\item Judge, T. A., Thoresen, C. J., Bono, J. E., \& Patton, G. K. 2001.
The job satisfaction\textendash{}job performance relationship: A qualitative
and quantitative review. \emph{Psychological Bulletin}, 127: 376-407. 
\item Kruschke, J. K., Aguinis, H., \& Joo, H. 2012. The time has come:
Bayesian methods for data analysis in the organizational sciences.
\emph{Organizational Research Methods}, 15: 722-752.
\item Lepak, D. P., Liao, H., Chung, Y., \& Harden, E. E. 2006. A conceptual
review of human resource management systems in strategic human resource
management research. \emph{Research in Personnel and Human Resources
Management}, 25: 217-271.
\item Morris, C. N. 1992. Hierarchical models for combining information
and for meta-analysis. \emph{Bayesian Statistics, }4: 321-44.
\item Oh, I. S., Wang, G., \& Mount, M. K. 2011. Validity of observer ratings
of the five-factor model of personality traits: A meta-analysis. \emph{Journal
of Applied Psychology}, 96: 762-773.
\item Paauwe, J., Wright, P. M., \& Guest, D. E. 2013. HRM and performance:
What do we know and where should we go? In J. Paauwe, D. E. Guest,
\& P. M. Wright (Eds), \emph{HRM \& performance: Achievements \& challenges}
(pp. 1-13). Chichester: Wiley. 
\item Pfeffer, J. \& Salancik, G. R. 1978. \emph{The external control of
organizations: A resource dependence perspective}. New York, NY: Harper
and Row. 
\item Podsakoff, P. M., MacKenzie, S. B., Lee, J. Y., \& Podsakoff, N. P.
2003. Common method biases in behavioral research: a critical review
of the literature and recommended remedies. \emph{Journal of Applied
Psychology}, 88: 879-903. 
\item Rogatko, A. 1992. Bayesian-approach for meta-analysis of controlled
clinical-trials. \emph{Communications in Statistics -- Theory and
Methods}, 21: 1441-1462. 
\item Rosenthal, R. 1991. \emph{Meta-analytic procedures for social research}.
Newbury Park, CA: Sage. 
\item Sin, H. P., Nahrgang, J. D., \& Morgeson, F. P. 2009. Understanding
why they don\textquoteright{}t see eye to eye: An examination of leader\textendash{}member
exchange (LMX) agreement. \emph{Journal of Applied Psychology}, 94:
1048-1057.
\item Smith, T. C., Spiegelhalter, D. J., Thomas, A. 1995. Bayesian approaches
to random-effects meta-analysis: A comparative study. \emph{Statistics
in Medicine, }14: 2685-2699. 
\item Spiegelhalter, D. J., Best, N. G., Carlin, B. P., \& Linde, A. v.
d. 2002. Bayesian measures of model complexity and fit. \emph{Journal
of the Royal Statistical Society, Series B (Statistical Methodology),
}64: 583-639.
\item Steel, P. D. G. \& Kammeyer-Mueller, J. 2008. Bayesian Variance Estimation
for Meta-Analysis: Quantifying Our Uncertainty. \emph{Organizational
Research Methods, }11: 54-78.
\item Subramony, M. 2009. A meta-analytic investigation of the relationship
between HRM bundles and firm performance. \emph{Human Resource Management},
48: 745-768.
\item Sutton, A. J. \& Abrams, K. R. 2001. Bayesian methods in meta-analysis
and evidence synthesis. \emph{Statistical Methods in Medical Research,
}10: 277-303.
\item Tukey, J. W. 1960. A survey of sampling from contaminated distributions.
In I. Olkin, J.G. Ghurye, W. Hoeffding, W. G. Madoo, \& H. Mann (Eds.),
\emph{Contributions to probability and statistics}. Stanford, CA:
Stanford University Press. 
\item Viswesvaran, C., \& Ones, D. S. 1995. Theory testing: Combining psychometric
meta-analysis and structural equations modeling. \emph{Personnel Psychology},
48: 865-885. 
\item Zhang, Z., Lai, K., Lu, Z., \& Tong, X. 2013. Bayesian inference and
application of robust growth curve models using student\textquoteright{}s
t distribution. \emph{Structural Equation Modeling, }20(1): 47-78.
\item Zyphur, M. J., \& Oswald, F. L. 2013. Bayesian estimation and inference:
A user\textquoteright{}s guide. \emph{Journal of Management}.
\end{reflist}

\section{APPENDIX A}

With the prior and the information from the first study, the posterior,
based on Bayes' Theorem, is
\begin{eqnarray*}
p(\zeta|z_{1}) & = & \frac{p(\zeta)p(z_{1}|\zeta)}{p(z_{1})}\\
 & = & \frac{\frac{1}{\sqrt{2\pi\psi_{0}}}\exp\left[-\frac{(\zeta-\zeta_{0})^{2}}{2\psi_{0}}\right]\frac{1}{\sqrt{2\pi\phi_{1}}}\exp\left[-\frac{(z_{1}-\zeta)^{2}}{2\phi_{1}}\right]}{p(z_{1})}\\
 & = & \frac{\frac{1}{\sqrt{2\pi\psi_{0}}}\frac{1}{\sqrt{2\pi\phi_{1}}}\exp\left[-(\frac{1}{2\psi_{0}}+\frac{1}{2\phi_{1}})\zeta^{2}+2(\frac{1}{2\psi_{0}}\zeta_{0}+\frac{1}{2\phi_{1}}z_{1})\zeta-(\frac{\zeta_{0}^{2}}{2\psi_{0}}+\frac{z_{1}^{2}}{2\phi_{1}})\right]}{p(z_{1})},\\
 & = & \frac{D\exp\left[-\frac{1}{2}(A\zeta^{2}+2B\zeta+C)\right]}{p(z_{1})}\\
 & = & \frac{D\exp\left[-\frac{(\zeta-\frac{B}{A})^{2}}{2\frac{1}{A}}-\frac{1}{2}(C-\frac{B^{2}}{A})\right]}{p(z_{1})}
\end{eqnarray*}
where 
\begin{eqnarray*}
A & = & \frac{1}{\psi_{0}}+\frac{1}{\phi_{1}}\\
B & = & \frac{1}{\psi_{0}}\zeta_{0}+\frac{1}{\phi_{1}}z_{1}.\\
C & = & \frac{\zeta_{0}^{2}}{\psi_{0}}+\frac{z_{1}^{2}}{\phi_{1}}
\end{eqnarray*}
The denominator is 
\begin{eqnarray*}
p(z_{1}) & = & \int_{-\infty}^{+\infty}(D\exp\left[-\frac{(\zeta-\frac{B}{A})^{2}}{2\frac{1}{A}}-\frac{1}{2}(C-\frac{B^{2}}{A})\right])d\zeta\\
 & = & D\exp\left[-\frac{1}{2}(C-\frac{B^{2}}{A})\right]\times\sqrt{2\pi\frac{1}{A}}
\end{eqnarray*}
Therefore, the posterior is 
\[
p(\zeta|z_{1})=\frac{1}{\sqrt{2\pi\frac{1}{A}}}\exp\left[-\frac{(\zeta-\frac{B}{A})^{2}}{2\frac{1}{A}}\right],
\]
a normal distribution with mean 
\begin{eqnarray*}
B/A & = & \frac{\frac{1}{\psi_{0}}\zeta_{0}+\frac{1}{\phi_{1}}z_{1}}{\frac{1}{\psi_{0}}+\frac{1}{\phi_{1}}}=\frac{\phi_{1}\zeta_{0}+\psi_{0}z_{1}}{\phi_{1}+\psi_{0}}
\end{eqnarray*}
and variance 
\[
1/A=\frac{1}{\frac{1}{\psi_{0}}+\frac{1}{\phi_{1}}}.
\]

\section{APPENDIX B}

With the power parameter $\alpha_{1}$, the posterior
\begin{eqnarray*}
p(\zeta|z_{1}) & = & \frac{p(\zeta)[p(z_{1}|\zeta)]^{\alpha_{1}}}{p(z_{1})}\\
 & = & \frac{\frac{1}{\sqrt{2\pi\psi_{0}}}\exp\left[-\frac{(\zeta-\zeta_{0})^{2}}{2\psi_{0}}\right]\left\{ \frac{1}{\sqrt{2\pi\phi_{1}}}\exp\left[-\frac{(z_{1}-\zeta)^{2}}{2\phi_{1}}\right]\right\} ^{\alpha_{1}}}{p(z_{1})}\\
 & = & \frac{\frac{1}{\sqrt{2\pi\psi_{0}}}\left(\frac{1}{\sqrt{2\pi\phi_{1}}}\right)^{\alpha_{1}}\exp\left[-\frac{(\zeta-\zeta_{0})^{2}}{2\psi_{0}}-\exp\left[-\frac{(z_{1}-\zeta)^{2}}{2\phi_{1}/\alpha_{1}}\right]\right]}{}\\
 & = & \frac{D\exp\left[-(\frac{1}{2\psi_{0}}+\frac{1}{2\phi_{1}^{*}})\zeta^{2}+2(\frac{1}{2\psi_{0}}\zeta_{0}+\frac{1}{2\phi_{1}^{*}}z_{1})\zeta-(\frac{\zeta_{0}^{2}}{2\psi_{0}}+\frac{z_{1}^{2}}{2\phi_{1}^{*}})\right]}{p(z_{1})},\\
 & = & \frac{D\exp\left[-\frac{1}{2}(A\zeta^{2}+2B\zeta-C)\right]}{p(z_{1})}\\
 & = & \frac{D\exp\left[-\frac{(\zeta-\frac{B}{A})^{2}}{2\frac{1}{A}}-\frac{1}{2}(C-\frac{B^{2}}{A})\right]}{p(z_{1})}
\end{eqnarray*}
where 
\begin{eqnarray*}
A & = & \frac{1}{\psi_{0}}+\frac{1}{\phi_{1}^{*}}\\
B & = & \frac{1}{\psi_{0}}\zeta_{0}+\frac{1}{\phi_{1}^{*}}z_{1},\\
C & = & \frac{\zeta_{0}^{2}}{\psi_{0}}+\frac{z_{1}^{2}}{\phi_{1}^{*}}
\end{eqnarray*}
and $\phi_{1}^{*}=\phi_{1}/\alpha_{1}$. From Appendix A, the posterior
is $N(B/A,1/A)$ where 
\begin{eqnarray*}
B/A & = & \frac{\frac{1}{\psi_{0}}\zeta_{0}+\frac{1}{\phi_{1}^{*}}z_{1}}{\frac{1}{\psi_{0}}+\frac{1}{\phi_{1}^{*}}}=\frac{\phi_{1}^{*}\zeta_{0}+\psi_{0}z_{1}}{\phi_{1}^{*}+\psi_{0}}=\frac{\frac{\phi_{1}}{\alpha_{1}}\phi_{1}+\psi_{0}z_{1}}{\frac{\phi_{1}}{\alpha_{1}}+\psi_{0}}\\
1/A & = & \frac{1}{\frac{1}{\psi_{0}}+\frac{1}{\phi_{1}^{*}}}=\frac{1}{\frac{1}{\psi_{0}}+\frac{\alpha_{1}}{\phi_{1}}}.
\end{eqnarray*}

\section{APPENDIX C}

Show the posterior
\begin{eqnarray*}
p(\zeta|z_{1},z_{2},\alpha_{1},\alpha_{2}) & = & \frac{p(\zeta|z_{1},\alpha_{1})[p(z_{2}|\zeta)]^{\alpha_{2}}}{p(z_{2})}\\
 & = & \frac{\frac{1}{\sqrt{2\pi\psi_{1}^{*}}}\exp\left[-\frac{(\zeta-\zeta_{1}^{*})^{2}}{2\psi_{1}^{*}}\right]\left\{ \frac{1}{\sqrt{2\pi\phi_{2}}}\exp\left[-\frac{(z_{2}-\zeta)^{2}}{2\phi_{2}}\right]\right\} ^{\alpha_{2}}}{p(z_{2})}\\
 & = & \frac{\frac{1}{\sqrt{2\pi\psi_{1}^{*}}}\left(\frac{1}{\sqrt{2\pi\phi_{2}}}\right)^{\alpha_{2}}\exp\left[-\frac{(\zeta-\zeta_{1}^{*})^{2}}{2\psi_{1}^{*}}-\exp\left[-\frac{(z_{1}-\zeta)^{2}}{2\phi_{1}/\alpha_{1}}\right]\right]}{p(z_{2})}\\
 & = & \frac{D\exp\left[-\frac{1}{2}(A\zeta^{2}+2B\zeta-C)\right]}{p(z_{1})}\\
 & = & \frac{D\exp\left[-\frac{(\zeta-\frac{B}{A})^{2}}{2\frac{1}{A}}-\frac{1}{2}(C-\frac{B^{2}}{A})\right]}{p(z_{1})}
\end{eqnarray*}

The denominator is 
\begin{eqnarray*}
p(z_{1}) & = & \int_{-\infty}^{+\infty}(D\exp\left[-\frac{(\zeta-\frac{B}{A})^{2}}{2\frac{1}{A}}-\frac{1}{2}(C-\frac{B^{2}}{A})\right])d\zeta\\
 & = & D\exp\left[-\frac{1}{2}(C-\frac{B^{2}}{A})\right]\times\sqrt{2\pi\frac{1}{A}}
\end{eqnarray*}

The posterior is $N(B/A,1/A)$ where 
\begin{eqnarray*}
B/A & = & \frac{\frac{1}{\psi_{1}^{*}}\zeta_{1}^{*}+\frac{1}{\phi_{2}^{*}}z_{2}}{\frac{1}{\psi_{1}^{*}}+\frac{1}{\phi_{2}^{*}}}=\frac{\frac{1}{\psi_{0}}\zeta_{0}+\frac{\alpha_{1}}{\phi_{1}}z_{1}+\frac{\alpha_{2}}{\phi_{2}}z_{2}}{\frac{1}{\psi_{0}}+\frac{\alpha_{1}}{\phi_{1}}+\frac{\alpha_{2}}{\phi_{2}}}\\
1/A & = & \frac{1}{\frac{1}{\psi_{1}^{*}}+\frac{1}{\phi_{1}^{*}}}=\frac{1}{\frac{1}{\psi_{0}}+\frac{\alpha_{1}}{\phi_{1}}+\frac{\alpha_{2}}{\phi_{2}}}
\end{eqnarray*}

\section{APPENDIX D}

The joint posterior distribution is
\begin{eqnarray*}
p(\zeta,\tau,\zeta_{i}|z_{i},\phi_{i},\alpha_{i}) & \propto & p(\zeta)p(\tau)\prod_{i=1}^{m}p_{\alpha_{i}}(z_{i},\zeta_{i}|\zeta,\tau)\\
 & = & p(\zeta)p(\tau)\prod_{i=1}^{m}\left[p_{\alpha_{i}}(z_{i}|\zeta_{i},\phi_{i})p(\zeta_{i}|\zeta,\tau)\right]\\
 & \propto & \frac{1}{\sqrt{2\pi\psi_{0}}}\exp\left[-\frac{(\zeta-\zeta_{0})^{2}}{2\psi_{0}}\right]\tau^{-\delta_{0}-1}\exp\left[-\frac{\gamma_{0}}{\tau}\right]\\
 &  & \times\left[\prod_{i=1}^{m}(2\pi\phi_{i})^{-\alpha_{i}/2}\right]\exp\left[-\sum_{i=1}^{m}\frac{(z_{i}-\zeta_{i})^{2}}{2\phi_{i}/\alpha_{i}}\right](2\pi\tau)^{-m/2}\exp\left[-\frac{\sum_{i=1}^{m}(\zeta_{i}-\zeta)^{2}}{2\tau}\right].
\end{eqnarray*}
Now we obtain the conditional posterior distributions. 

First, we get the conditional posterior distribution of $\tau$, which
is
\[
p(\tau|\zeta_{i},\zeta,\alpha_{i})\propto\tau^{-\delta_{0}-1-m/2}\exp\left[-\frac{2\gamma_{0}+\sum(\zeta_{i}-\zeta)^{2}}{2\tau}\right].
\]
Therefore, the posterior is inverse Gamma distribution IG($\delta_{0}+m/2,\gamma_{0}+[\sum(\zeta_{i}-\zeta)^{2}]/2$).

Second, the conditional posterior distribution of $\zeta$ is 
\[
p(\zeta|\zeta_{i},\tau)\propto\exp\left[-\frac{(\zeta-\zeta_{0})^{2}}{2\psi_{0}}-\frac{\sum(\zeta_{i}-\zeta)^{2}}{2\tau}\right].
\]
Therefore, the conditional posterior is a normal distribution
\[
N\left(\frac{\frac{\sum_{i=1}^{m}\zeta_{i}}{\tau}+\frac{\zeta_{0}}{\psi_{0}}}{\frac{m}{\tau}+\frac{1}{\psi_{0}}},\frac{1}{\frac{m}{\tau}+\frac{1}{\psi_{0}}}\right).
\]

Third, the conditional posterior distribution of $\zeta_{i}$ is
\[
p(\zeta_{i}|\zeta,z_{i},\tau,\alpha_{i})\propto\exp\left[-\frac{(z_{i}-\zeta_{i})^{2}}{2\phi_{i}/\alpha_{i}}-\frac{(\zeta_{i}-\zeta)^{2}}{2\tau}\right],
\]
which is a normal distribution
\[
N\left(\frac{\frac{\zeta}{\tau}+\frac{z_{i}\alpha_{i}}{\phi_{i}}}{\frac{1}{\tau}+\frac{\alpha_{i}}{\phi_{i}}},\frac{1}{\frac{1}{\tau}+\frac{\alpha_{i}}{\phi_{i}}}\right).
\]

\section{APPENDIX E}

The joint posterior distribution for the meta-regression model is
\begin{eqnarray*}
p(\bm{\beta},\tau|z_{i},\zeta_{i},\alpha_{i}) & \propto & p(\bm{\beta})p(\tau)\prod p_{\alpha_{i}}(z_{i},\zeta_{i}|\bm{\beta},\tau)\\
 & = & p(\bm{\beta})p(\tau)\prod p_{\alpha_{i}}(z_{i}|\zeta_{i},\phi_{i})p(\zeta_{i}|\bm{\beta},\tau)\\
 & \propto & |\mathbf{\Psi}_{0}|^{-1/2}\exp\left[-\frac{1}{2}(\bm{\beta}-\bm{\zeta}_{0})'\mathbf{\Psi_{0}}^{-1}(\bm{\beta}-\bm{\zeta}_{0})\right]\tau^{-\delta_{0}-1}\exp\left[-\frac{\gamma_{0}}{\tau}\right]\\
 &  & \times\prod\left\{ (2\pi\phi_{i})^{-\alpha_{i}/2}\exp\left[-\sum\frac{(z_{i}-\zeta_{i})^{2}}{2\phi_{i}/\alpha_{i}}\right](2\pi\tau)^{-m/2}\exp\left[-\frac{\sum(\zeta_{i}-\mathbf{x}_{i}\bm{\beta})^{2}}{2\tau}\right]\right\} .
\end{eqnarray*}
The conditional posterior distribution of $\tau$ is 
\begin{equation}
p(\tau|\bm{\beta},\zeta_{i})\propto\tau^{-(\delta_{0}+m/2)-1}\exp\Bigg[-\frac{\gamma_{0}+\sum(\zeta_{i}-\mathbf{x}_{i}\bm{\beta})^{2}/2}{\tau}\Bigg]
\end{equation}
Therefore, $\tau|\bm{\beta},\zeta_{i}\sim IG(\delta_{0}+m/2,\gamma_{0}+\sum(\zeta_{i}-\mathbf{x}_{i}\bm{\beta})^{2}/2)$.
The conditional posterior distribution for $\bm{\beta}$ is

\begin{equation}
p(\bm{\beta}|\tau,\zeta_{i})\propto\exp\Bigg[-\frac{1}{2}(\bm{\beta}-\bm{\zeta}_{0})'\mathbf{\Psi}_{0}^{-1}(\bm{\beta}-\bm{\zeta}_{0})\Bigg]\exp\Bigg[\frac{\sum(\zeta_{i}-\mathbf{x}_{i}\bm{\beta})^{2}}{2\tau}\Bigg].\label{beta-1}
\end{equation}
Let $\bm{\zeta}=(\zeta_{1},\zeta_{2},\cdots,\zeta_{m})'$ be the vector
of $\zeta_{i}'s$, and $\hat{\beta}$ be the least square estimate
such that $\hat{\bm{\beta}}=(\mathbf{X}'\mathbf{X})^{-1}\mathbf{X}'\bm{\zeta}$
with $\mathbf{X}=(\mathbf{x}_{1},\ldots,\mathbf{x}_{m})'$ as the
design matrix. Then the conditional posterior distribution of $\beta$
is a multivariate normal distribution 
\[
N\Bigg(\bigg(\mathbf{\Psi}_{0}^{-1}+\frac{\mathbf{X}'\mathbf{X}}{\tau}\bigg)^{-1}\bigg(\mathbf{\Psi}_{0}^{-1}\bm{\zeta}_{0}+\frac{\mathbf{X}'\mathbf{X}}{\tau}\hat{\bm{\beta}}\bigg),(\mathbf{\Psi}_{0}^{-1}+\frac{\mathbf{X}'\mathbf{X}}{\tau})^{-1}\Bigg)
\]
For $\zeta_{i}$, its conditional distribution is 
\[
p(\zeta_{i}|\beta,\tau)\propto\exp\Bigg[-\frac{(z_{i}-\zeta_{i})^{2}}{2\phi_{i}/\alpha_{i}}\Bigg]\exp\Bigg[-\frac{(\zeta_{i}-X_{i}\beta)^{2}}{2\tau}\Bigg],
\]
a normal distribution 
\[
N(\frac{\frac{\alpha_{i}z_{i}}{\phi_{i}}+\frac{X_{i}\bm{\beta}}{\tau}}{\frac{\alpha_{i}}{\phi_{i}}+\frac{1}{\tau}},\frac{1}{\frac{\alpha_{i}}{\phi_{i}}+\frac{1}{\tau}}).
\]

\section{FOOTNOTES}
\begin{APAenumerate}
\item The results are based on a total of 10,000 iterations with the first
4,000 iterations as burn-in.
\item {*} $p<0.05$.
\item Power 1: each study is given a power of 1. Power 2: the reliability
of financial performance is used as power. Power 3: the two studies
with the largest sample sizes are given a power of 0.1. Power 4: studies
with correlations larger than 0.2 are given a power of 0.5, otherwise,
1.\end{APAenumerate}

\end{document}